\documentclass[epj]{svjour}
\usepackage{latexsym}
\usepackage[dvips]{graphicx}

\begin{document}
\title{Shot-noise statistics in diffusive conductors}
\author{P.-E.~Roche\inst{1} \and B.~Dou\c{c}ot\inst{1,2}}
\institute{Laboratoire de Physique de la Mati\`ere Condens\'ee, Ecole Normale Sup\'erieure, 24 rue Lhomond, 75231 Paris Cedex 05, France
\and Laboratoire de Physique Th\'{e}orique et Hautes Energies, CNRS UMR 7589, UniversitŽ Paris VII - Denis Diderot, 4 place Jussieu, 75252 Paris Cedex 05, France
}
\date{Received: date / Revised version: date}
%
\abstract{
We study the full probability distribution of the charge transmitted through a mesoscopic diffusive conductor during a measurement time $\Delta$.
We have considered a semi-classical model, with an exclusion principle in a discretized single-particle phase-space. In the large $\Delta$ limit,
numerical simulations show a universal probability distribution which agrees very well with the quantum mechanical prediction of Lee, Levitov
and Yakovets [PRB \textbf{51} 4079 (1995)] for the charge counting statistics. Special attention is given to its third cumulant, including an
analysis of finite size effects and of some experimental constraints for its accurate measurement.
} 
\maketitle
\section{Introduction}
\label{intro}

The electrical current across a conductor results from a flow of charge carriers. Their discreetness is responsible
for current fluctuations $\delta i$ -called shot noise- characterized by a noise power $S_{I}=2\int_{}^{}\overline{\delta i(t)\delta i(O)} dt$
proportional to the mean current $I$. For example, a current I of uncorrelated charge e will exhibit a full
shot noise $S_{I}=2.e.I$. For a recent review on shot noise and for general references, see~\cite{Blanter2000}.

In diffusive conductors shorter that the electron-phonon mean free length, an universal shot noise power $S_{I}=(2.e.I) / 3$
has been predicted~\cite{Beenakker92,deJong92,Nazarov94,Altshuler,Blanter97,Nagaev,deJong95,Sukhorukov} and validated 
experimentally~\cite{Liefrink,Steinbach,Schoelkopf,Henny}.

This $1/3$ reduction over the full shot noise focused a lot of attention because
it has been derived in two different frameworks: quantum mechanics (scattering matrix theory~\cite{Beenakker92,deJong92,Nazarov94}
and Green function technique~\cite{Altshuler,Blanter97}) and semi-classical mechanics (Boltzmann-Langevin approach including Pauli's exclusion
principle~\cite{Nagaev,deJong95,Sukhorukov}). While Landauer saw a numerical coincidence between these results~\cite{Landauer}, 
de Jong and Beenakker~\cite{deJongPHYSA96} minimized the role of phase coherence, which entails that an exclusion principle is the only
irreducible concept behind the two treatments.

To be more complete, we should also mention the studies of the shot noise reduction in non-degenerate
diffusive conductors:~\cite{Gonzales,Beenakker99,Schomerus}.

The shot noise \textit{power} can be seen as the statistical variance of the measurable current fluctuations.
(by \textit{measurable current}, we mean: current averaged over a time scale much larger than
the electron-electron correlation time, which is always the case in electrical set-ups, even for very highpass systems.)
In addition to the variance, the shot noise
should also be characterized by all the higher moments of current fluctuations.
In this paper, we will focus on the information beyond shot noise power, in the perspective of testing
the merits of classical versus quantum mechanics in the noise reduction process.

The first significant result beyond the shot noise power was derived by Levitov and Lesovik~\cite{Levitov}.
Starting from the description of a charge-counting operator, they derived the full statistics of transmitted charges
through a point contact, or in other words the probability that Q charges are transmitted
through a lumped scatterer during a given measurement duration $\Delta$ that is large compared to the
electronic correlation times. Note that in this large $\Delta$ limit, $Q/\Delta$ become what we called a ÒmeasurableÓ current.
Two years later, Lee et al.~\cite{Lee} combined this result with the bimodal distribution of the transmission eigenvalues
predicted by the random matrix theory for a phase coherent diffusive conductor~\cite{Dorokhov}. In the large $\Delta$ limit,
they obtained the statistics $P_{q}(Q)$ for the transmitted charges $Q$ by this fully quantum derivation.

In this paper, we will show that the full statistics of transmitted charge through a diffusive conductor
is a semi-classical quantity. To do so, we show that the prediction of Lee et al. can be recovered in a semi-classical
model which accounts for an exclusion principle but not for phase coherence. This result generalizes to diffusive
conductors an earlier derivation for double-barrier junctions~\cite{deJongPRB96}.

\section{The model}
\label{model}

A minimal modeling of out-of-equilibrium semi-classical mesoscopic conductor consists in a 1D open system with a flow and a back-flow of charges of
opposite velocities. This degenerate model can be easily adapted to a one dimensional chain: each site being associated with a back-scattering probability
and a Pauli exclusion rule. Such degenerate systems are often considered in the theoretical physics literature as "simple exclusion process" 
or traffic models.
Indeed some of the analytical
results on shot noise power mentioned previously have been derived with such models~\cite{deJong95,Liu}. In order to strictly overlap with these previous
studies, we stick with the exact sequence of dynamic rules considered by Liu, Eastman and Yamamoto~\cite{Liu}.

The model consists in a chain of N sites. Each site is either empty, occupied by a right or a left-moving charge or by two charges propagating in
opposite directions. During each time step, all charges are shifted to the next site, according to their direction. After this, each charge is likely
to back-scatter with a probability $r$, provided that the resulting state is empty. The system is maintained out of equilibrium by dissymmetrical boundary
conditions: charges are injected at each time step at one end of the chain while both ends act as perfect absorber for charges incoming from the chain.

For the model parameters considered in this paper, over $10^{8}$ samples are sometimes required for the statistics to converge up to the desired accuracy. In practice, the numerical
simulation could still be performed on a desktop computer thanks to a code which core is restricted to low level-processor-instructions.
Basically, each configuration of the chain occupation can be coded by the binary representation of two integers, one for each charge direction.
Each time step is a combination of a register shift (right or left), a bit increment to account for the charge
injection at the boundary, and a bit-to-bit comparison of the two integers to check for the scattering which are compatible
with the Pauli exclusion principle. The effectiveness of scattering is set by precomputed series of random bits,
refreshed when necessary. A special attention was dedicated to the validation of the random numbers algorithm. In addition to this direct simulation,
two semi-analytical methods provided a cross validation of the results
up to N=6 sites. For longer chains, those methods were either too CPU-time or memory consuming.

\begin{figure}
\centerline{\includegraphics[width=8cm]{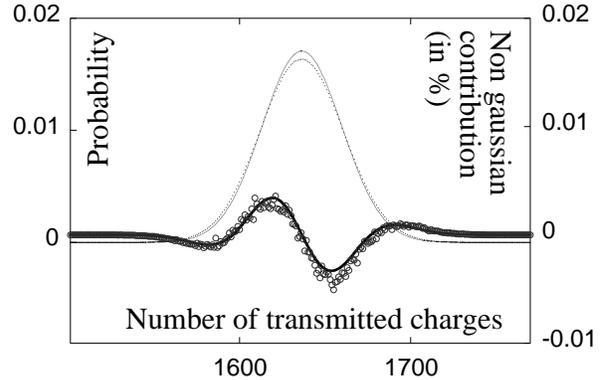}}
\caption{\emph{Left scale}: probability distribution from the number of transmitted charges during $\Delta=18000$ time units.
Each site presents a $r=50\%$ back-scattering probability, which results in a $G=\langle Q_{\Delta}\rangle /\Delta=9\%$ conductivity.
Dots: present semi-classical model. Thin line: Lee et al. quantum model for the same average transmission~\cite{Lee}.
\emph{Right scale}: difference between each distribution and their gaussian fit. Open circles: present model, thick line: Lee et al. model.}
\label{fig:1}       
\end{figure}

\section{Results}
The dots on Figure 1 present the probability distribution of the transmitted charges $Q_{\Delta}$ for typical parameters of the
semi-classical model (measurement duration: $\Delta=18000$ time units~\cite{buffer}, site's back-scattering probability: $r=50\%$).
For comparison, the distribution $Q_{q}$ predicted by Lee et al. quantum model is plotted for the same average transmission
(thin line)~\cite{Lee}.

The central limit theorem states that both these distributions converge to a gaussian distribution when $\Delta \rightarrow \infty$.
In this large $\Delta$ limit and within finite size corrections, it has been demonstrated that the distributions variances
of both models -i.e. the shot noise power- become equal~\cite{deJong95,Lee}.
More precisely, in the large $\Delta$ limit, de Jong and Beenakker showed that when the number of sites $N$ increases, the Fano 
factor $\langle (Q_{\Delta}-\langle Q_{\Delta}\rangle )^{2}\rangle / \langle Q_{\Delta}\rangle $ converges to $\langle (Q_{q}-\langle Q_{q}\rangle )^{2}\rangle / \langle Q_{q}\rangle =1/3$.

Beyond the gaussian approximation, the non-gaussian contribution of both distributions is also plotted on figure 1 (right scale).
The central finding of this work is the excellent agreement between the semi-classical (open circles) and 
the quantum models (thick line) which strongly suggests that a semi-classical picture fully accounts for the
whole statistics of transmitted charges in a diffusive conductor.

\section{Finite size effects: the signature of correlations}
\label{correlations}

The remaining of this paper will focus on the finite size effects associated with the three parameters of the model: the chain
length or number of sites $N$, the measurement duration $\Delta$ and the back-scattering probability $r$ on each site.

The third cumulant $\langle (Q_{\Delta}-\langle Q_{\Delta}\rangle )^{3}\rangle $ provides a useful measure of the
deviation from the gaussian distribution and it can be directly compared with the quantum model 
prediction $\langle (Q_{q}-\langle Q_{q}\rangle )^{3}\rangle =1/15.\langle Q_{q}\rangle$~\cite{Lee}.
In this last relation, the linear scaling with $\langle Q_{q}\rangle $ is simply a consequence of the 
fact that for large measurement duration ($\Delta=\infty$ in this model), $\langle Q_{q}\rangle $ is the sum of almost
uncorrelated random variables which effective number scales linearly with $\Delta$. One of the motivation for considering 
cumulants is precisely
that they behave linearly with respect to the addition of independent variables. Therefore, the physics of the short
time scales electron correlations is captured by the numerical prefactor $1/15$. 
In the following, it will be useful to call the ratio of the third cumulant by the average transmission the 
\textit{third Fano factor} $F_{3}$, in reference to the usual Fano factor $F$.

\begin{figure}
\centerline{\includegraphics[width=8cm]{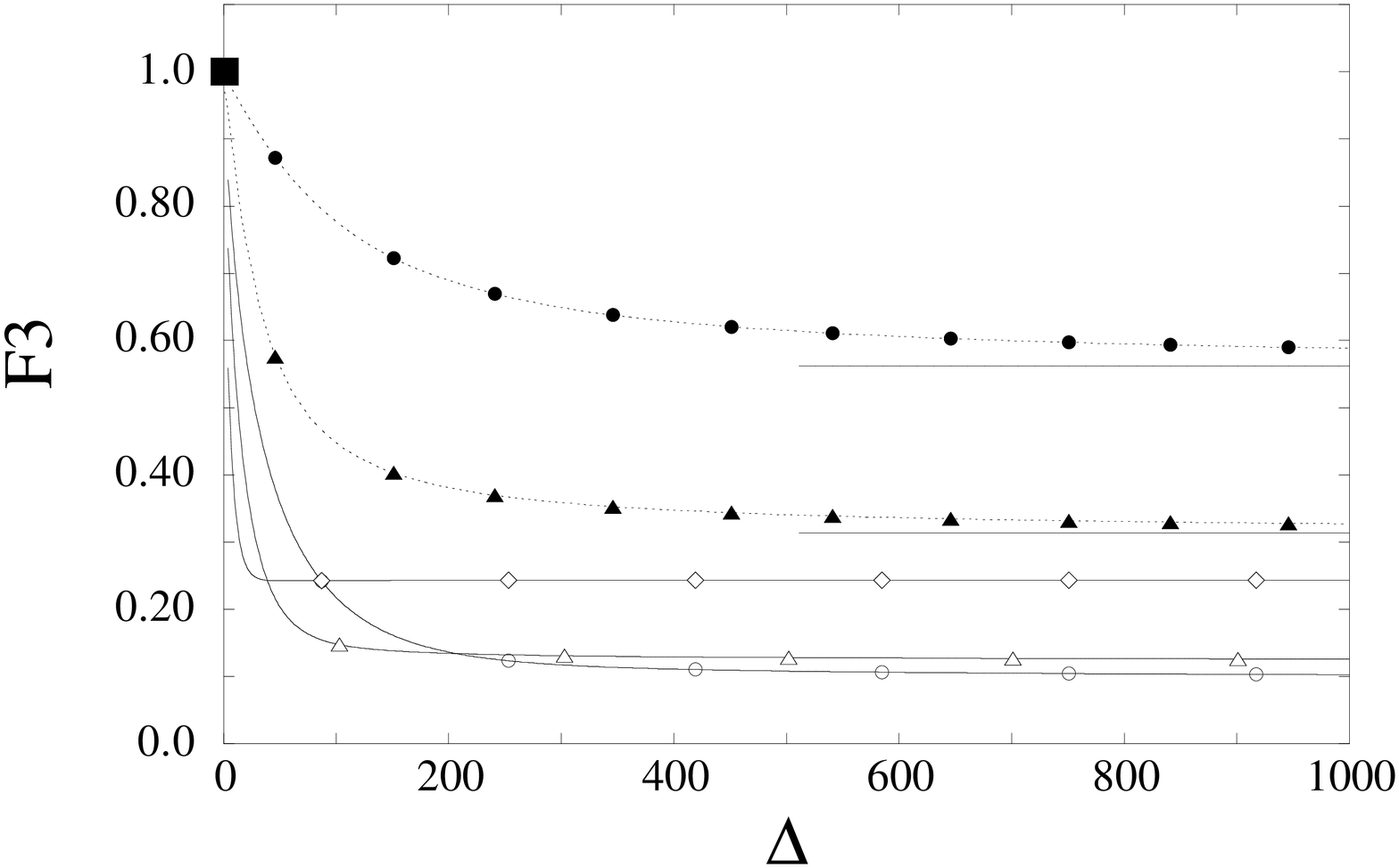}}
\caption{Dependence of the third Fano factor $F_{3}=\langle (Q_{\Delta}-\langle Q_{\Delta}\rangle )^{3}\rangle /\langle Q_{\Delta}\rangle$ with the
measurement duration $\Delta$ for the present model (solid lines) and for the same model without exclusion principle (dashed line).
For all curves the back-scattering probability is $r=85\%$. The number of sites per chain is $\diamondsuit$: 2,
triangles: 4, circles: 7. The \textit{transient} regime evidences an anti-correlation time between the charge arrivals.}
\label{fig:2}
\end{figure}

Figure 2 presents the semiclassical third Fano factor $F_{3}=\langle (Q_{\Delta}-\langle Q_{\Delta}\rangle )^{3}\rangle /\langle Q_{\Delta}\rangle$
versus the measurement duration $\Delta$ for different combinations of $N$ and $r$ (solid lines). After
a \textit{transient} regime, the signal settles to a constant level. The transient regime duration defines a correlation time
$\tau$. Beyond this regime, for $\Delta >> \tau$, the random variable $Q_{\Delta}$ scales like $\Delta / \tau$ as expected and
the third Fano factor tends towards a constant value. Note that the measurement duration
$\Delta=18000$ used in figure 1 -to compare $Q_{\Delta}$ and $Q_{q}$- is validated by a plot similar to figure 2 for $r=50\%$.

The simulations show that $\tau$ increases when the chain length is increased as one would expect for
a similar model without exclusion principle. In this latter case, we shall show that $\tau$ is of the order of the 
scattering time through the chain. Let us compare further the model
with and without exclusion principle.
On figure 2, $F_{3}$ versus $\Delta$ is plotted for this second model for the same back-scattering probability and
chain lengths (dashed lines). The \textit{transient} regimes are significantly larger in this second model. In the $\Delta=0$
limit, $F_{3}=1$ as expected from a Poisson distribution while for large $\Delta$, $F_{3}$ converges to $(1-G).(1-2G)$ where
$G=\langle Q_{\Delta}\rangle /\Delta$ is the conductivity. More extensive calculations confirmed this large $\Delta$ limit,
which correspond to a binomial distribution.

This result can be understood with the basic sketch of figure 3. To begin with, the main feature of the model in the absence of exclusion principle
is that electrons emitted at different times are completely uncorrelated. The probability distribution of the transmitted charge is therefore
completely determined by the probability distribution for a single particle to leave the system on the right, as a function of the time spent
inside it. Let us denote by $\tau$ the width of this distribution. The left picture on figure 3 illustrates what happens as $\Delta \gg \tau$.
Most detected particles have been emitted in region 2. For them, the measurement window covers the whole span of likely values for their
lifetime in the system, before leaving on the right. Region 1 corresponds to particles emitted early, so they have to spend more time than on
average in the system in order to contribute to the signal. Symmetrically, region 3 corresponds to particles spending a shorter time than
on average in the system. Region 1 and 3 have a width of order $\tau$, whereas region 2 lasts for a time equivalent to $\Delta$ as $\Delta \gg \tau$.
All the particles emitted in region 2 contribute independently to the signal with the same probability $P_{\infty}$, so the measured charge
follows a binomial distribution: 

\begin{displaymath}
\mathcal{P}(Q_{\Delta}) \simeq \frac{\Delta !}{{Q_{\Delta}}! {( \Delta - Q_{\Delta} )}!}.(P_{\infty})^{Q_{\Delta}}.(1-P_{\infty})^{\Delta -Q_{\Delta}}
\end{displaymath}

The right picture on figure 3 illustrates the opposite situation, when $\Delta \ll \tau$. First, we expect anticorrelations between charges
measured in the time intervals $[t,t+\Delta ]$ and $[t',t'+\Delta ]$ if $\Delta <\mid t-t'\mid \ll \tau$, simply because both measurements
involve particles which have been emitted in a same time interval of width proportional to $\tau$, and the same particle cannot be detected twice.
As $\Delta \ll \tau$, the probability for each particle to contribute to the signal becomes very small, so we get a Poisson distribution.
The cross-over from this Poisson distribution to the binomial one occurs for $\Delta \sim \tau$, and is precisely due to the correlations
between the small time intervals $[t,t+\Delta]$ and $[t',t'+\Delta]$ when $\Delta <\mid t-t' \mid \ll \tau$. The main value of this overly
simple model is to illustrate the role of the correlation time $\tau$. It also emphasizes the need to give an accurate modelling of the
injection process. The binomial law for $\Delta \gg \tau$ is strongly connected to the fact that the particles are injected periodically
in the system. If the injection process is instead taken to follow a Poisson distribution, the scale $\tau$ doesn't appear and the transmitted
charge obeys a Poisson distribution, for all values of $\Delta$. These two features (role of $\tau$ and importance of the injection mechanism)
are of course expected to be found in more complex and more realistic models, such as the model simulated here, although full analytical
treatments are not easily available. Let us now return to the case with the exclusion principle.

\begin{figure}
\centerline{\includegraphics[width=8cm]{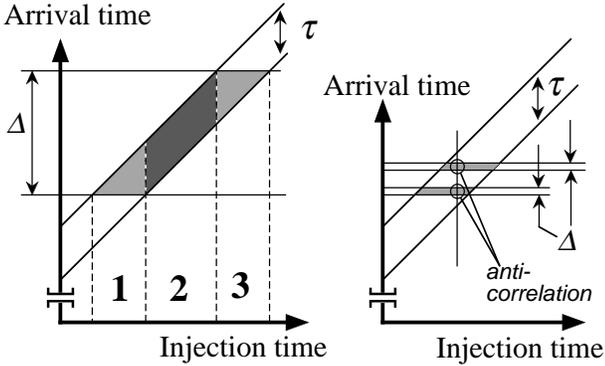}}
\caption{A simple way to analyze the model in the absence of exclusion principle. Electrons emitted at different times are uncorrelated.
They have a finite probability to contribute to the signal if their typical arrival time (lying in the tilted band of slope unity and width $\tau$)
lies in the measurement window (the horizontal band of width $\Delta$). Left large $\Delta$ limit ($\Delta \gg \tau$). Right: small $\Delta$
limit ($\Delta \ll \tau$). The picture shows two different measurement windows separated by an interval shorter than $\tau$, which leads to strong
anticorrelations between the two signals.}
\label{fig:3}
\end{figure}

\begin{figure}
\centerline{\includegraphics[width=8cm]{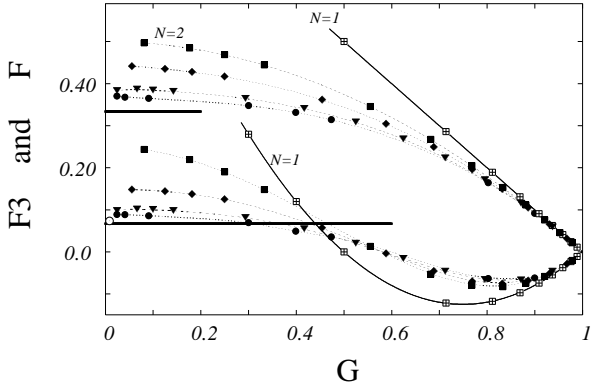}}
\caption{Dependence of the Fano $F$ and third Fano factors $F_{3}$ versus the conductivity $G=\langle Q_{\Delta}\rangle / \Delta $.
The symbols are the data for N=1, 2, 3, 6, 10, 28 sites ; the dotted lines through those symbols are guideline.
The thin lines are the dependences for a lumped scatterer predicted by the coherent scattering formalism: $(1-G)$ and $(1-G).(1-2G)$.
The two plateaux at $F=1/3$ and $F_{3}=1/15$ are predicted by Lee et al.'s quantum derivation for a diffusive conductor.}
\label{fig:4}
\end{figure}

Figure 4 presents the Fano factor $F$ and the third Fano $F_{3}$ factor versus the conductance $G=\langle Q_{\Delta}\rangle /\Delta$ for
various numbers of sites N. This plot completes a similar one published in~\cite{Liu} for the Fano factor. The continuous lines are the
factors $F=(1-G)$ and $F_{3}=(1-G).(1-2G)$ for a point scatterer predicted by the coherent scattering formalism~\cite{Lee}.
The agreement with the
data in the $G\rightarrow 1$ limit can be understood easily: for $N.r<1$, the mean free path becomes longer than
the chain and the whole system can be considered as a lumped (or point) scatterer. On contrary, in the low conductivity limit, each
charge undergoes many collisions in the chain and a diffusive-like behavior is expected. It is known that as the chain length increases,
the data for the Fano factor converge to a plateau at $1/3$~\cite{deJong95,Lee}. Figure 4 shows that the
same type of asymptotic plateau emerges for the third Fano factor data. 
A plateau at $F_{3}=1/15$, plotted in figure 4, is compatible with the data.

\begin{figure}
\centerline{\includegraphics[width=8cm]{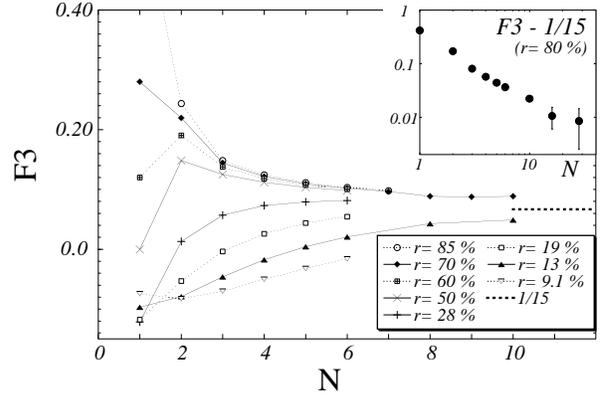}}
\caption{Dependence of the third Fano $F_{3}$ factors with the number of sites $N$.
The plateau at $F_{3}=1/15$ is predicted by Lee et al. \emph{Inset}: $F_{3}-1/15$ versus $N$ for $r=80\%$ up to $N=28$.}
\label{fig:5}
\end{figure}

Another point of view on these data supports the conjecture of an asymptotic plateau at $1/15$. Figure 5
presents the third Fano $F_{3}$ factor versus the number of sites $N$ for various back-scattering probabilities $r$. The residual discrepancy
between the $1/15$ limit and the $N=10$ data appears as a finite size effect on $N$. The inset shows the difference
between the third Fano factor and $1/15$ as a function of the number of sites N for $r=80\%$. It is interesting to note
that the convergence is compatible with a power law with a $(-1)$ exponent, which may be related to long range correlations 
between charge carriers induced by the exclusion principle.

\section{Perspective for experiments}
To the knowledge of the authors, experiments on the statistics of the transmitted charge 
have never been extended beyond the shot noise power. This may not be surprising since such experiments are difficult for a 
fundamental reason: the physical phenomena revealed by higher cumulants are related to the correlation time scale $\tau$ while the
experimental probes (amperemetre, \ldots) always have a time constant $\Delta$ such that $\Delta \gg \tau$. As we recalled earlier,
in this limit the statistics of transmitted charges are very close to a gaussian and a large number $n$ of statistical samples
are required to discern the non-gaussian contributions, such as the third Fano factor $F_{3}$. In the following, we show that
under specific experimental conditions, these contributions remain measurable.

We shall first consider the experimental analog of the cumulants $\langle (Q_{\Delta}-\langle Q_{\Delta}\rangle )^{2}\rangle$
and $\langle (Q_{\Delta}-\langle Q_{\Delta}\rangle )^{3}\rangle$ from which $F$ and $F_{3}$ are derived. 
A central point is the relation between the number of transmitted charges $Q_{\Delta}$ and the currents $i$
measured by the amperemetre ; our analysis differs from a previous one on this point~\cite{Levitov2001}. The dynamical response of 
the amperemetre is characterized by a cut-off frequency $f_{B}$ or the corresponding time scale $\Delta = 1/f_{B}$ which can be seen
as the duration of the \textit{charge counting} from which \textit{each} current output is inferred.
Consequently, the relation between $Q_{\Delta}$ and each current output $i$ should then be:

\begin{displaymath}
Q_{\Delta}\sim \Delta .i\sim i/f_{B}
\end{displaymath}

At this point, two comments are necessary. Firstly, $f_{B}$ should be understood as the effective frequency at the amperemetre output,
which is set by the whole set-up, including filters, cables,\ldots Secondly, in practice, the measurement chain
is rarely a low pass but rather a bandpass, in order to filter out the low frequency noise or as a consequence of the typical
specifications of RF elements. In the following, $f_{B}$ should be understood as the measurement bandwidth rather than the cut-off frequency.
Note that we are implicitly assuming that the shot noise is white, which is reasonable for the experimental conditions considered below.

Using the previous equation, the Fano factors $F$ and $F_{3}$ can be written as :

\begin{eqnarray*}
\overline{(i-\overline{i})^3} & = & F_{3}.\overline{i}.(e.f_{B})^2
\end{eqnarray*}

where the overline denotes the average over the bandwidth $f_{B}$.
The experimental uncertainty on $F_{3}$ is mostly due to the
uncertainty on $\overline{(i-\overline{i})^3}$, which results itself from undesirable current noises $i_{N}$ in the circuit
(such as amplifier noise and thermal noises) and from the finiteness of the number of statistical samples $n$.
Assuming that $i$ and $i_{N}$ are independent and quasi-gaussian,
the resulting \textit{signal to noise} $S/N$ of $\overline{(i-\overline{i})^3}$ and $F_{3}$ is indeed:

\begin{eqnarray*}
S/N & = & \overline{(i-\overline{i})^3} / \sqrt{var((i+i_{N})^3)} \\
    & = & \overline{(i-\overline{i})^3} / \sqrt{6.(\overline{(i-\overline{i})^2}+\overline{i_{N}^2})^3/n}
\end{eqnarray*}

where we used the relation between the variance $var$ of the statistical estimate of the third cumulant of a gaussian and its second cumulant.
To a reasonable approximation for our purpose, $\overline{(i-\overline{i})^2}$ is the sum of the shot noise $2F.e.\overline{i}$ and the
thermal noise $4k.T/R$ across the conductor of resistance $R$:

\begin{displaymath}
\overline{(i-\overline{i})^2}\sim (2F.e.\overline{i}+4k.T/R).f_{B}/2
\end{displaymath}

If $S_{N}$ is the noise power density of $i_{N}$, we also have:

\begin{displaymath}
\overline{i_{N}^2}=S_{N}.f_{B}/2
\end{displaymath}

The four last equations can be combined to give $n$ as a function of the other parameters including $S/N$ and $\overline i$.
If the amperemetre output is sampled at a rate $f_{s}$ (oversampling being excluded) and if we set $S/N=10$, the duration $T$
of the experiment will be:

\begin{displaymath}
T=\frac {n}{f_{s}} \sim 75.\frac{(2F.e.\overline{i}+4k.T/R+S_{N})^3}{f_{s}.f_{B}.\overline{i}^2.e^4.F_{3}^2}
\end{displaymath}

For $f_{B}=f_{s}=1 GHz$, $S_{N}=10^{-24} A^2/Hz$, $T=4 K$, $R=50 \Omega $ and $\overline {i}=10-1000 \mu A$, we find 
an experiment duration $T<1 min$, comfortable for an experimentalist.

The typical RF frequencies that we find for $f_{B}$ has consequences on the procedure for processing the
amperemetre output~\cite{amperemetre}. A calibrated analogic device can certainly perform the $i\to i^3$ function but even if a fully 
analog signal processing is feasible~\cite{analog}, we believe that a numerical 
acquisition is preferable here. Given the typical parameters $f_{s}$ and $T$, real time
storing of all the data is not possible with present technologies but a histogram of the measured current is enough
for a post-processing of the cumulants and real time histograms can be acquired at frequencies up to several GHz.
This signal processing procedure would also allow to extract more 
information than just $F_{3}$. In particular, figure 1 shows that the deviation from the gaussian has a $\sim \sim$ shape
which is nearly orthogonal to the gaussian and not efficiently sensed by the $i\to i^3$ projection function of the third cumulant.
Since this $\sim \sim$ function is known theoretically, it could be used as a projection vector for the experimental histograms.
It is clear that such a processing would significantly increases the $S/N$ ratio.

\section{Concluding remarks}
\label{consequences}

The main result of this work is the perfect quantitative agreement between the classical model (with an exclusion principle in phase space)
and the full quantum treatment of reference ~\cite{Lee} for the complete probability distribution of transmitted charge through a 1D
diffusive conductor, during a measurement time $\Delta$. Such an equivalence has already been discussed by many researchers in the last decade,
either at the level of the second moment of this distribution (the noise power)~\cite{Nagaev,deJong95,Sukhorukov} or for the special case
of a double-barrier system~\cite{deJongPRB96}. However, our impression is that a first principle understanding of why this should be true is
still lacking. Indeed, an advantage of the direct numerical simulation we have performed is to yield to the complete stationary out of
equilibrium probability distribution on the configuration space of our system, which is the analogue of the Sinai-Ruelle-Bowen
measure for continuous dynamical systems~\cite{Dorfman}. This information goes in principle much beyond the one contained in the
Boltzmann stationary distribution or even the more refined Boltzmann-Langevin approach to the fluctuations in the single particle
distribution around its stationary value. The latter formalism has proved very effective for the computation of noise reduction factors in
various models~\cite{Nagaev,deJong95,Sukhorukov}. But it relies on a key assumption on the fluctuations of the density current in single
particle phase-space, namely it is a Poissonian random process with local correlations in space and time. Although this is a very
reasonable assumption, as argued in~\cite{Kogan}, a rigorous derivation of this behavior requires in principle the exact knowledge of two
particle correlations from the stationary distribution in configuration space. This goes far beyond the knowledge of the Boltzmann distribution.
A natural question arising from the present work is to compare these two particle correlations obtained in the numerical simulation with the
assumption made in~\cite{Kogan}. We hope to get a clear answer on this important point in the near future.
In fact, the strong similarity between our model and the simple exclusion process considered in~\cite{Eyink} suggests that the
stationary distribution in configuration space is a very complex object. Recently, Derrida et al have uncovered some fascinating
properties of the rare large fluctuations of such distributions, which exhibit a strongly non-local character~\cite{Derrida}. It
would be very interesting to analyze the stationary distributions obtained here along these lines, although it may be impossible
to derive exacts analytical expressions as for the simple exclusion model.

Finally, it would be very useful to generalize the Boltzmann-Langevin approach of~\cite{Nagaev,deJong95,Sukhorukov} to the computation
of the complete distribution of the transmitted charge. At this point, we may object that a full quantum-mechanical derivation
is already available~\cite{Lee}. however, this derivation is involving a quenched averaging over the possible impurity configurations.
For a given mesoscopic sample, this may be justified by an ergodic hypothesis, so that averaging the transmission matrix over a finite
energy window $e.V$ becomes equivalent to impurity averaging. But this may break down for very small systems. Furthermore, semi-classical
models are very flexible, since they have been extended to treat interaction effects and inelastic processes
~\cite{Gonzales,Beenakker99,Nagaev95,Kozub}. Again, we believe a detailed analysis of the stationary distribution in configuration
space is required to make further progress.

We want to thank O. Verzelen, T. Jolicoeur, G.ÊBastard and R.ÊFerreira for sharing their computing resources. We are
also grateful to N.ÊRegnault for some programmer tricks, to H. Willaime and P. Tabeling for technical support with the
preliminary experiment mention in~\cite{amperemetre} and to H. Bouchiat, S. Gu\'eron and B. Reulet for their feed-back.


\begin{thebibliography}{}
\bibitem{buffer}
On a 64 bit processor, if the chain is represented by N=28 bits allocated in the middle of the 64 bits word, that leaves 18 unused bits
on each side. Those bits can be used as buffers, to differ and speed up the counting of transmitted and reflected charges, every 18
time steps. This very 18 number explains the somehow odd measurement durations $\Delta$ that are chosen in our computation.
\bibitem{amperemetre}
In this paper, \textit{Amperemetre} should be understood as a generic name for a detector. For example, we verified experimentally 
that using of 3 detectors (voltmeters) connected across the conductor under study, we could decrease the uncertainty on $F_{3}$ below the
level expected if we had used one single detector.
This performance was possible because the third cumulant $\overline{(R.i-\overline{R.i})^3}$ has been calculated
from the output of the 3 detectors according to $\overline{(R.i_{1}-\overline{R.i_{1}}).(R.i_{2}-\overline{R.i_{2}}).(R.i_{3}-\overline{R.i_{3}})}$,
which allows to significantly decrease the effect of the voltmeters' voltage noise. For high resistance conductor, a similar trick could
decrease the nuisance from the current noise of the detectors. Note also that this multi-detectors technique can be generalized for higher
moments.
\bibitem{analog}
In this case $f_{s}$ is the cut-off frequency of the set-up.
\end{thebibliography}
\end{document}